\documentclass[letterpaper,twocolumn,12pt]{article}
\usepackage{graphicx}
\usepackage{jpc}

\setkeys{Gin}{width=7.75cm,clip=true}

\makeatletter
  \newcommand{\ru}{}%
  \newcommand{\onlinecite}[1]{{\let\@cite\ru(\cite{#1})}\ru}%
  \renewcommand\@cite[1]{\textsuperscript{#1}}

  \renewcommand\section{
    \@startsection{section}{1}{0cm}{0.5cm plus1ex minus .2ex}
    {0.3cm plus1ex minus.2ex}{\reset@font\bf}
  }%
\makeatother

\begin{document}
\bibliographystyle{jpc}

\title{\textbf{
Towards a density functional description of liquid pH$_{\mathbf 2}$ 
}}

\author{
\textbf{J. Navarro\footnote{IFIC}, F. Ancilotto\footnote{Universit\`a di Padova.},}\\
\textbf{M. Barranco\footnote{ 
To whom correspondence should be addressed.
E-mail: manuel@ecm.ub.es. Phone: +34 93 402 1184. Fax: +34 93 402 1198}, 
M. Pi\footnote{Universitat de Barcelona}}}

\onecolumn
\maketitle

\begin{quote}
IFIC (CSIC and Universidad de Valencia). Apdo. 22085, 46071 Valencia, Spain.\\
Dipartimento di Fisica, `G.Galilei', Universit\`a di Padova, via
Marzolo 8, I-35131 Padova, Italy and
CNR-IOM-Democritos, I-34014 Trieste, Italy. \\
Departament ECM, Facultat de F\'{\i}sica,
and IN$^2$UB, Universitat de Barcelona. Diagonal 647, 08028 Barcelona, Spain.
\end{quote}

\begin{abstract}

A finite-temperature density functional approach to describe the properties of parahydrogen in the liquid-vapor coexistence region is presented. The first proposed functional 
is zero-range, where the density-gradient term is adjusted so as to reproduce the
surface tension of the liquid-vapor interface at low temperature. 
The second functional is finite-range and, while it is fitted to reproduce bulk pH$_2$ properties only,
it is shown to yield surface properties in good agreement with
experiments. These functionals are used to study the surface thickness of the liquid-vapor
interface, the wetting transition of parahydrogen on a planar Rb 
model surface, and homogeneous cavitation in bulk liquid pH$_2$.
\\


\end{abstract}

\newpage
%\twocolumn

\section{Introduction}

Under normal conditions of pressure and temperature hydrogen is a gas formed by H$_2$ molecules. 
The molecule exists in two isomeric forms, para (pH$_2$) and ortho (oH$_2$), 
which differ in the coupling of their nuclear spins: antiparallel ($J=0$) and parallel ($J=1$), 
respectively. At low temperatures, near the boiling point (20.3~K), 
hydrogen can be catalyzed in nearly all parahydrogen.
%hydrogen is nearly all parahydrogen.

The purpose of this paper is to present a free energy density functional for liquid parahydrogen.
Our approach is largely inspired by density functional methods currently employed to study liquid 
helium at very low temperature.\cite{Str87,Dal95,Bar06}
Indeed, the pH$_2$ molecule and the $^4$He atom have in common that both are spinless bosons 
subject to a weak van der Waals interaction.
While helium remains liquid down to $T=0$ K, the stable low-temperature bulk phase of hydrogen,
in spite of the lighter mass of the pH$_2$ molecule, is an hpc solid. This is a consequence 
of the attractive interaction between two pH$_2$ molecules, which is a factor of four stronger than 
the He-He one. The density functional for parahydrogen presented here
is valid in the temperature range 14 K $\leq T \leq $ 32 K, {\it i.e.},
between the triple (13.8~K) and nearly the critical point (32.938~K).
While the functional is still reliable in the neighborhood of the liquid-vapor coexistence region, 
the extrapolation at lower temperatures to study, for instance, the metastable overcooled liquid, 
still remains an open question.

Ginzburg and Sobyanin\cite{Gin72} have suggested that liquid parahydrogen could exhibit superfluidity, analogously to liquid helium, below some critical temperature that they estimated to be at around 6~K. 
To have superfluid pH$_2$ it is therefore necessary to bring the liquid below its saturated vapor pressure curve. Attempts to produce  a superfluid by supercooling the normal liquid below the
triple point have been insofar unsuccessful.\cite{Mar83} As a possible way to stabilize the liquid
phase of pH$_2$ at low temperatures several authors have considered restricted geometries in order 
to reduce the effective attraction between molecules. Indeed, the lowering of the melting point 
compared to the bulk liquid is a well-known and rather general phenomenon in clusters,
see {\em e.g.} Ref. \onlinecite{Alo05}.
Path integral Monte Carlo (PIMC) simulations by Sindzingre {\em et al.}\cite{Sin91} predicted that 
(pH$_2$)$_N$ clusters with $N=13$ and 18 molecules are superfluid below about 2~K. 
This prediction motivated two experiments\cite{Gre00,Tej04,Mon09} and a much larger number of 
theoretical studies on small (pH$_2$)$_N$ clusters, see {\em e.g.} 
Refs. \onlinecite{War10,Alo10,Gua10} and references therein.
Concerning two-dimensional  pH$_2$ systems, PIMC calculations of films have found that introducing
some alkali metal atoms stabilizes a liquid hydrogen phase which undergoes a superfluid 
transition below $\approx 1.2$~K.\cite{Gor97} Moreover, experiments on adsorption of H$_2$ films on
alkali metals substrates\cite{Mis94,Ros98} have reported the existence of a wetting transition, 
analogous to what is known for helium.

A density functional (DF) for liquid parahydrogen at finite temperatures could help to 
describe these phenomena. Indeed, DF methods have become increasingly popular in recent years as  
useful computational tools to study the properties of classical and quantum inhomogeneous 
fluids,\cite{Eva89} especially for large systems for which these
methods provide a good compromise between accuracy and computational
cost, yielding results in agreement with experiment or with more microscopic approaches. 

A first step in that direction was made by Boninsegni and Szybisz,\cite{Bon04} who
constructed a density functional to study the low temperature energetics and structural behavior of 
H$_2$ films adsorbed on Cs and Li substrates. They fitted the DF parameters by using as 
input experimental informations from the solid phase, such as the ground state 
equilibrium density, the sublimation energy, and the bulk modulus. 
Our approach is based on a different strategy, as its aim is to provide
a DF that can be reliably used in the liquid-vapor coexistence region.
 
This paper is organized as follows. In Sec. 2 we describe the density functional for liquid
parahydrogen in two versions, neglecting or not finite-range effects.  
The validity of the DF is assessed in Sec. 3 by comparing the calculated surface tension 
with experiment. The DF is thus used to study the wetting of a Rb planar surface by
pH$_2$, and the results are compared to those obtained by Quantum Monte Carlo simulations. 
As a final application, in Sec. 4 we describe homogeneous cavitation in liquid pH$_2$.
A brief summary and outlook are presented in Sec. 5.

\section{The density functional}

Following the procedure previously employed for liquid $^4$He, in particular the works of 
Guirao {\it et al.},\cite{Gui92} Dalfovo {\it et al.},\cite{Dal95} and Ancilotto {\it et al.},\cite{Anc00}
we present first a density functional for bulk liquid parahydrogen.
The extension to inhomogeneous liquids will be considered in the next section.
In the most general form, the free energy of a Bose system is assumed to be a DF,
which we write as

\begin{equation}
F = \int d{\bf r} \left\{ f_{ni}(\rho,T) + f_c(\rho,T) \right\} \, ,
\label{funci}
\end{equation}
where $f_{ni}$ is the free energy density for an ideal Bose gas, and
$f_c$ is the correlation energy density, which incorporates the dynamic correlations induced by the
interaction. The number of pH$_2$ molecules per unit volume is denoted by $\rho$.

\subsection{The non-interacting part}
The expression for $f_{ni}(\rho,T)$ at $T \ne 0$ is given in textbooks. 
Following Ref. \onlinecite{Hua87} it can be written as

\begin{equation}
f_{ni}=\rho \,k_BT \,ln(z)-{\frac{k_BT}{\lambda ^3}}\,g_{5/2}(z) \; ,
\label{eq:fni}
\end{equation}
where 

\begin{equation}
\lambda \equiv \sqrt{2\pi \hbar^2/m k_BT}
\end{equation}
is the pH$_2$ thermal wavelength, $m$ is the H$_2$ mass, and the fugacity $z$ is defined as:

\begin{equation}
z = \left\{
\begin{array}{ll}
1 & \mbox{   if $\rho \lambda ^3 \geq g_{3/2}(1) $ } \\
z_0 & \mbox{   if $\rho \lambda ^3 < g_{3/2}(1) $ } \; ,
\end{array}
\right.
\end{equation}
where $z_0$ is the root of the equation $\rho \lambda ^3=g_{3/2}(z)$. In the
above equations, $g_{p}(z)\equiv \sum _{l=1}^\infty z^l/l^p$. In the following we 
shall use units such that the Boltzmann constant $k_B$ is one, so that energy 
is measured in K. In these units $\hbar^2/m=$ 24.0626 K \AA$^{2}$. 

\subsection{The correlation part}

Once the functional Eq. (\ref{funci}) is defined, a complete thermodynamical description of
the bulk liquid can be achieved. In particular, the chemical potential and the pressure
are given by

\begin{equation}
\mu(\rho,T) =  \mu_{ni}(\rho,T) + \left( \frac{\partial f_c}{\partial \rho} \right)_T 
\end{equation}
and

\begin{equation}
p(\rho,T) = p_{ni}(\rho,T) + \rho^2 \left( \frac{\partial (f_c/\rho)}{\partial \rho} \right)_T \, .
\end{equation}
In the next subsections we propose a phenomenological expression for the correlation energy density,
inspired by previous works on liquid helium, which contains a few adjustable parameters
which are fixed so as to reproduce selected experimental properties of the liquid 
from the data collected by McCarthy {\em et al.}\cite{Mcc81} or the more recent compilation by 
Leachman {\em et al.}\cite{Lea09}
In fact, the latter compilation has been summarized in terms of a rather complex parametrization of the bulk liquid pH$_2$ free energy density which reproduces satisfactorily the available experimental data 
between the triple and the critical points in a wide range of pressure values. 

We present here a simpler yet less accurate functional that has the advantage of being easier to
generalize to finite systems (drops, liquid pH$_2$ in confined geometries) and presents
a more conventional separation into kinetic and potential terms. This may be of interest
for other applications, such as computing the excitation spectrum  as well as
the static and dynamic responses of these systems.\cite{Bar06} Besides, the present DF could 
in principle be extrapolated to lower temperatures near $T=0$, one of our medium-term goals.

\subsubsection{Zero-range}

We consider a polynomial in powers of the density 
to represent the correlation part of the free-energy density

\begin{equation}
f_c(\rho,T) = a_1(T) \rho^2 + a_2(T) \rho^3 + a_3(T) \rho^4 + a_4(T) \rho^5 \, .
\label{zero-r}
\end{equation}
For a given value of $T$, there are thus four free parameters. 
With this form, other quantities such as the chemical potential $\mu$, the pressure $p$, 
and the speed of sound $c$, can be written as
\begin{eqnarray}
\mu  &=& \mu_{ni} + 2 a_1 \rho + 3 a_2 \rho^2 + 4 a_3 \rho^3 + 5 a_4 \rho^4 
\nonumber \\
p  &=& p_{ni} + a_1 \rho^2 + 2 a_2 \rho^3 + 3 a_3 \rho^4 + 4 a_4 \rho^5 
\nonumber \\
m c^2 &=& \rho \frac{\partial \mu_{ni}}{\partial \rho} +
 2 a_1 \rho + 6 a_2 \rho^2 + 12 a_3 \rho^3 + 20 a_4 \rho^4 \, ,
\nonumber
\end{eqnarray}
where, for the sake of simplicity, we have omitted the $T$-dependence in the $a_i$ parameters.

These parameters are fixed by the following strategy.
At a given $T$, we determine the liquid-vapor equilibrium by imposing the equality 
between pressures and chemical potentials.
The liquid and vapor densities are experimentally known.\cite{Lea09}
This provides two conditions. The other two conditions are the fitting of the pressure and
speed of sound of the liquid phase at the saturated vapor pressure
to the corresponding experimental values.\cite{Lea09}
The resulting $a_i(T)$ parameters are given in Table \ref{coefs}.  
We shall refer to them as the DFpH2 set.

\begin{figure}[t]
\centerline{\includegraphics[width=14cm]{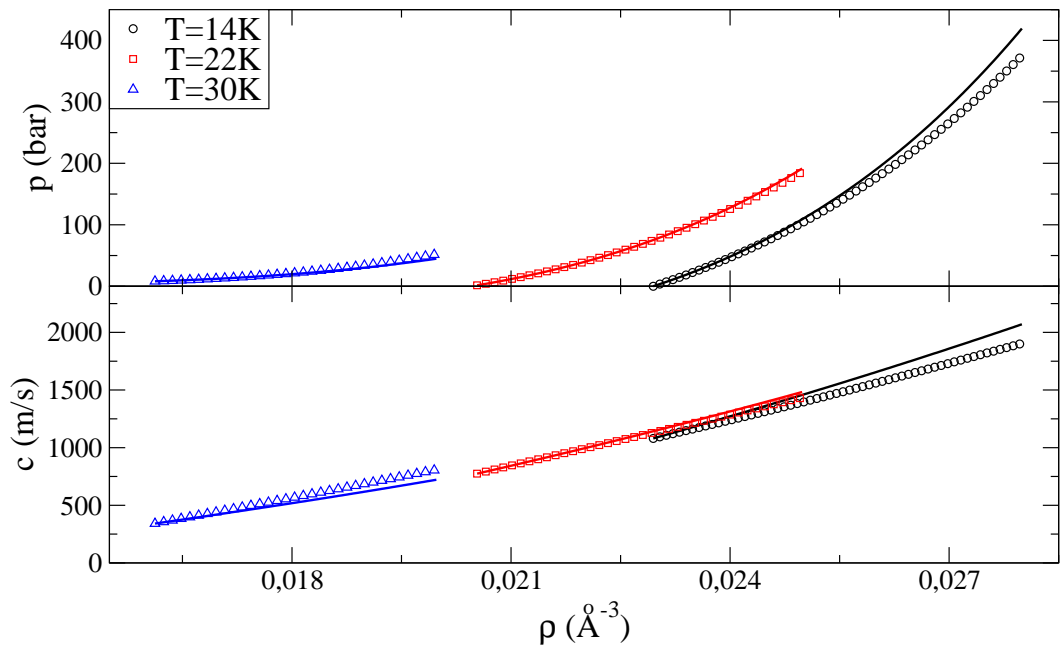}}
\caption{(Color online) Pressure and speed of sound: A comparison between 
experimental data (symbols) and DFpH2 results (lines) at three values of $T$.}
\label{fig1}
\end{figure}

In Fig.~\ref{fig1} the experimental and calculated pressure and speed of sound as a function of $\rho$ for three values of $T$ are compared. For each $T$, the first density value shown is that of the liquid at the liquid-vapor coexistence, where $p$ and $c$ coincide by construction with the experimental 
values.
It is worthwhile seeing that the DFpH2 parametrization can be safely used for 
densities well above those of the liquid at the saturation vapor pressure.

\subsubsection{Finite-range}

The functional Eq. (\ref{zero-r}) is a zero-range one, and it cannot account for inhomogeneities or
finite size effects induced by strong density compression. 
Inspired in the way this problem has been addressed in the case of liquid $^4$He,\cite{Dal95}
we derive from Eq.~(\ref{zero-r}) a finite-range density functional with the following prescriptions.
On one hand, the term in $\rho^2$ is replaced with 

\begin{equation}
a_1(T) \rho^2 \longrightarrow \frac{1}{2} \int {\rm d}{\bf r'} \rho({\bf r}) V(|{\bf r}-{\bf r'}|) 
\rho({\bf r'}) \; ,
\end{equation}
where
$V$ represents the H$_2$-H$_2$ pair interaction screened at short distances

\begin{equation}
V(r) = 4 \epsilon \left\{ \left(\frac{\sigma}{r}\right)^{12} - \left(\frac{\sigma}{r}\right)^6 \right\}
\quad  \quad {\rm if} \ r \ge h(T) \, ,
\end{equation}
and $V(r)=0$ otherwise. We have used the Lennard-Jones parameters
$\epsilon=34$ K and $\sigma=3.06$ \AA{}, which are slightly different from the
standard ones (34.2 and 2.96, respectively). The value of $h(T)$ is fixed by
the condition

\begin{equation}
a_1(T) = \frac{1}{2} \int {\rm d}{\bf r} \, V(r)  \; .
\end{equation}
The values of $h(T)$ are given in the last column of Table \ref{coefs}. 

On the other hand, a coarse-grained density is introduced in the remaining terms
of the correlation part Eq.~(\ref{zero-r}):

\begin{equation}
 a_i(T) \rho^{1+i} \longrightarrow a_i(T) \rho {\bar \rho}^i \quad , \quad i \ge 2
\; ,
\end{equation}
where

\begin{equation}
{\bar \rho}({\bf r}) = \int {\rm d}{\bf r'} w(|{\bf r}-{\bf r'}|) \rho({\bf r'})
\label{coarse}
\end{equation}
with

\begin{equation}
w(r) = \frac{3}{4 \pi \hat{h}^3(T)} \quad  \quad {\rm if} \ r \ge \hat{h}(T) \, ,
\end{equation}
and $w(r)=0$ otherwise.
For the reasons explained in Sec. 3.2, we have taken $\hat{h}(T)=1.14 \,h(T)$.

\section{Surface and interface properties}

Once the DF parameters have been adjusted to reproduce the bulk behavior of liquid pH$_2$ at 
finite temperatures, the predicted surface and interface properties provide a significant test 
about its validity. We present in this section the results of a series of calculations aimed at showing that 
our proposed DF can indeed be reliably used to study finite temperature 
properties of inhomogeneous pH$_2$ systems, such as a free-standing 
fluid-vapor interface or liquid pH$_2$ adsorbed on a solid surface.

We have first computed the liquid-vapor surface tension of pH$_2$ as a function of temperature,
and compared our results with the experimental data of Ref. \onlinecite{Mcc81}.
We obtain a liquid-vapor interface by minimizing the DF with the appropriate boundary conditions,
corresponding to a planar liquid pH$_2$ film in equilibrium with its own vapor. In
this case the solution depends only on the coordinate $z$ normal to the
surface, and provides the equilibrium density profile $\rho (z)$. From the knowledge of 
$\rho (z)$ one can extract values for the surface tension $\gamma(T)$. 
We distinguish in the following two different approaches used to compute this quantity, 
the first one being based on the zero-range DF described in Sec. 2.2.1, and the second one on the more accurate but computationally more expensive non-local, finite-range DF
described in Sec. 2.2.2.

\subsection{The pH$_2$ surface tension from zero-range DF calculations}

Our starting point is the zero-range free energy density functional

\begin{equation}
f=f_{ni}+f_{c} +\beta \, {\hbar ^2 \over 2 m}
\frac{(\nabla \rho)^2}{\rho} +\xi \, (\nabla \rho)^2 \; .
\label{dfzero}
\end{equation}
This is the simplest extension of Eq. (\ref{funci}) in Sec. 2 including the first
two terms in a gradient expansion to account for density inhomogeneities.
We make here the usual choice\cite{Dal95} $\beta = 1/4$ for the kinetic energy term,
while we fix $\xi$ to the value $\xi =19306.74$ K \AA$^5$, which allows to reproduce the 
experimental surface tension at $T=14$ K.\cite{Mcc81}

With such a form for the free energy density, the surface 
tension turns out to be\cite{Bar90}

\begin{equation}
\gamma(T) = 2\int _{\rho _{v}}^{\rho _{l}} \, {\rm d} \rho \,
[f(\rho )-f(\rho _{v}) -\mu (\rho -\rho _{v})
]^{1/2}\left(\frac{\beta}{\rho}+\xi\right)^{1/2} \; ,
\label{eq:sigma}
\end{equation}
$\rho _{l},\rho _{v}$ being the asymptotic (constant) density values
on the liquid and vapor side of the interface, respectively.

\begin{figure}[t]
\centerline{\includegraphics[width=14cm,clip]{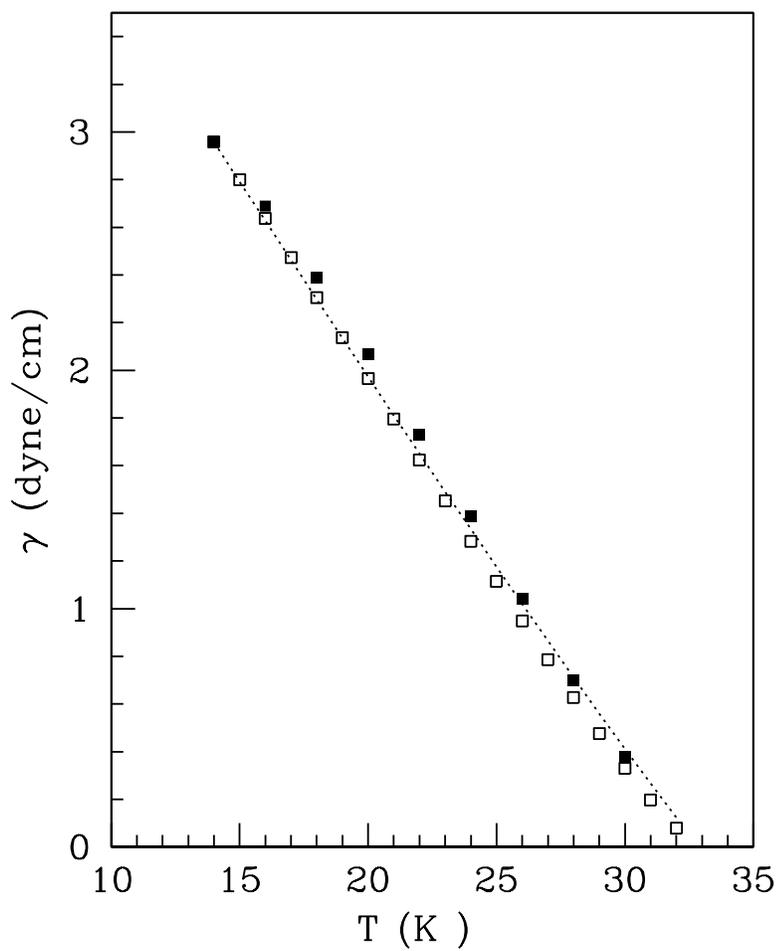}}
\caption{Calculated surface tension. Open squares: results of the
zero-range DF. Full squares: results of the finite-range DF. Dotted line: 
experimental data from Ref. \onlinecite{Mcc81}.
}
\label{fig2}
\end{figure}

We show in Fig. \ref{fig2} our results for the pH$_2$ surface tension.
The open squares correspond to the calculated points, while the dotted line
shows the experimental result from Ref. \onlinecite{Mcc81}.
It appears that, once the DF parameter $\xi $ is fixed to reproduce
the value at a given temperature, say $T=14$ K, the predictions for $\gamma(T)$
are quite accurate in the whole range of temperatures between the
triple and near the critical point.

\subsection{The pH$_2$ surface tension from finite-range DF
calculations}

We use here the finite-range description of the free energy 
density functional of pH$_2$, as discussed in Section 2.2.2.
Concerning the coarse-grained density defined in Eq. (\ref{coarse}), 
the empirical prescription $\hat {h}(T)=1.14\,h(T)$ is
the one which gives optimum agreement between the calculated 
and measured values for $\gamma(T)$.

We use a slab geometry in our calculations, with periodic boundary conditions 
whereby a thick planar liquid film is in contact with a vapor region.
Our calculations, which are effectively one-dimensional,
are performed within a region of length $z_{m}$.
Rather large values of $z_{m}$ and of the film thickness $t_{f}$ 
(no to be confused with the thickness of the liquid-vapor interface)
are necessary for well converged calculations (typically 
$z_{m}\sim 200$ \AA{} and $t_{f}\sim 150$ \AA{}), in order to let the system
spontaneously reach (i) the bulk liquid density in the interior of the film
and (ii) the bulk vapor density in the region outside the film.

The equilibrium density profile $\rho (z)$ of the fluid at the interface
is determined from the variational minimization of the grand potential,
leading to the following Euler-Lagrange (EL) equation:

\begin{equation}
\left\{-\frac{\hbar ^{2}}{2 m}
\frac{d^{2}}{dz^{2}}+U[\rho(z)]\right\}\sqrt{\rho(z)}=\mu \sqrt{\rho (z)} \; ,  
\label{eq:kseq}
\end{equation}
where the effective potential $U$ is defined as the functional
derivative of the free energy density,\cite{Dal95}
and the value of the chemical potential $\mu$ is fixed by the
normalization condition to the total number of H$_2$ molecules 
per unit area, $ \int dz\rho (z)=N/A $.

\begin{figure}[t]
\centerline{\includegraphics[width=14cm,clip]{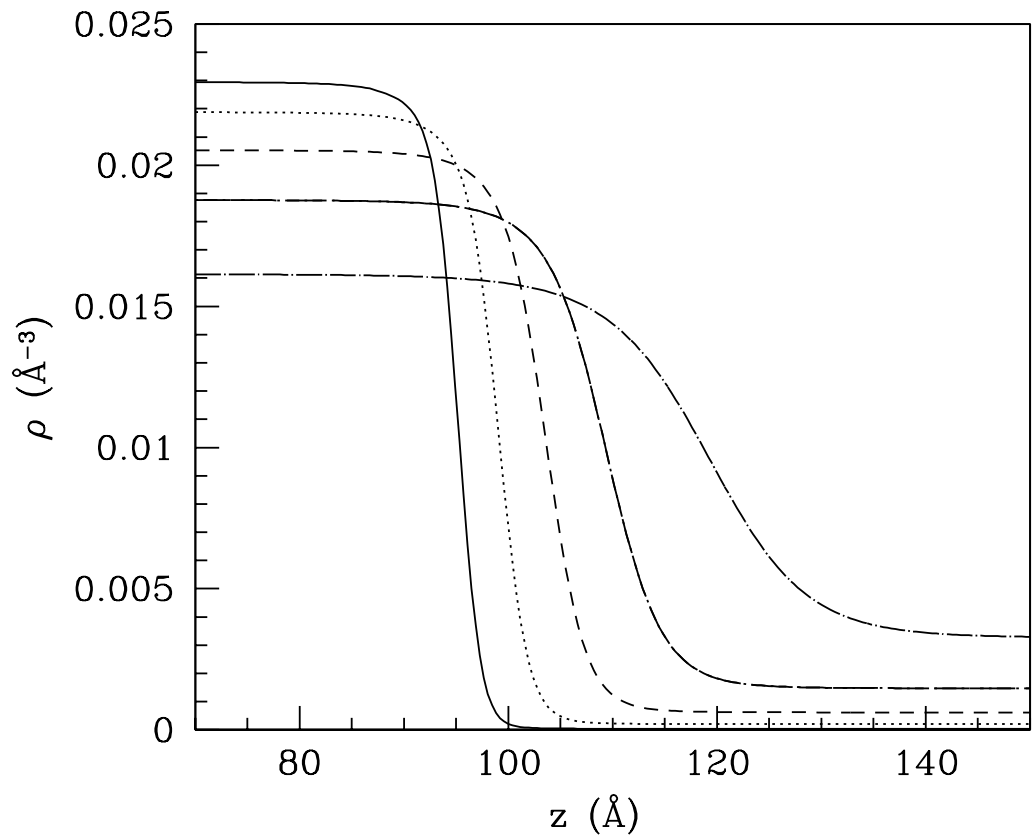}}
\caption{
Density profiles around the liquid-vapor interface calculated using the
finite-range DF. From top to bottom {\it in the liquid regime}, the
profiles correspond to $T=$ 14, 18, 22, 26, and 30 K.
The origin of the $z$ axis is arbitrary.
}
\label{fig3}
\end{figure}

In Fig. \ref{fig3} we show some selected equilibrium density
profiles around the liquid-vapor interface calculated using the
finite-range DF. As expected, the width of the interface 
increases monotonically with temperature. 
Density profiles for H$_2$ have been obtained in Ref. \onlinecite{Zha04}
using the Silvera-Goldman
potential\cite{Sil78} within a molecular simulation of the liquid-vapor
interface.
However, the rather thin liquid slab used there is probably
insufficient
to properly describe the liquid-vapor interface.
For this reason we do not attempt a quantitative comparison
between our calculated profiles and those reported in 
Ref. \onlinecite{Zha04}.

The liquid-vapor surface tension $\gamma$ can be directly computed
from the definition $\gamma = (\Omega +pV)/A$
once the equilibrium density profiles have been obtained. In the
previous expression, $\Omega=F-\mu N$ is the grand potential,
$p$ is the saturated vapor pressure at temperature $T$, 
$V$ is the volume of the system, and $A$ is the surface area. For the
one-dimensional problem considered here, one has:

\begin{equation}
\gamma = \frac{1}{2} \, \left[\int _0 ^{z_m} f[\rho(z)] {\rm d}z - \mu \int _0 ^{z_m}
\rho(z) {\rm d}z
+p\,z_m \right]  \; .
\label{eq:sigma_1d}
\end{equation}
The pressure $p$ can be conveniently calculated from the definition $p=[\mu
\rho _{v}-f(\rho _{v})]=[\mu \rho _{l}-f(\rho _{l})]$.
A factor $1/2$ appears in the previous equation to account for
the two free surfaces delimiting the liquid film in our slab geometry
calculations.
The surface tension obtained from the finite-range DF is also
shown in Fig. \ref{fig2} as full squares. The overall agreement with
the available experimental data\cite{Mcc81}
is quite satisfactory.

\begin{figure}[t]
\centerline{\includegraphics[width=14cm,clip]{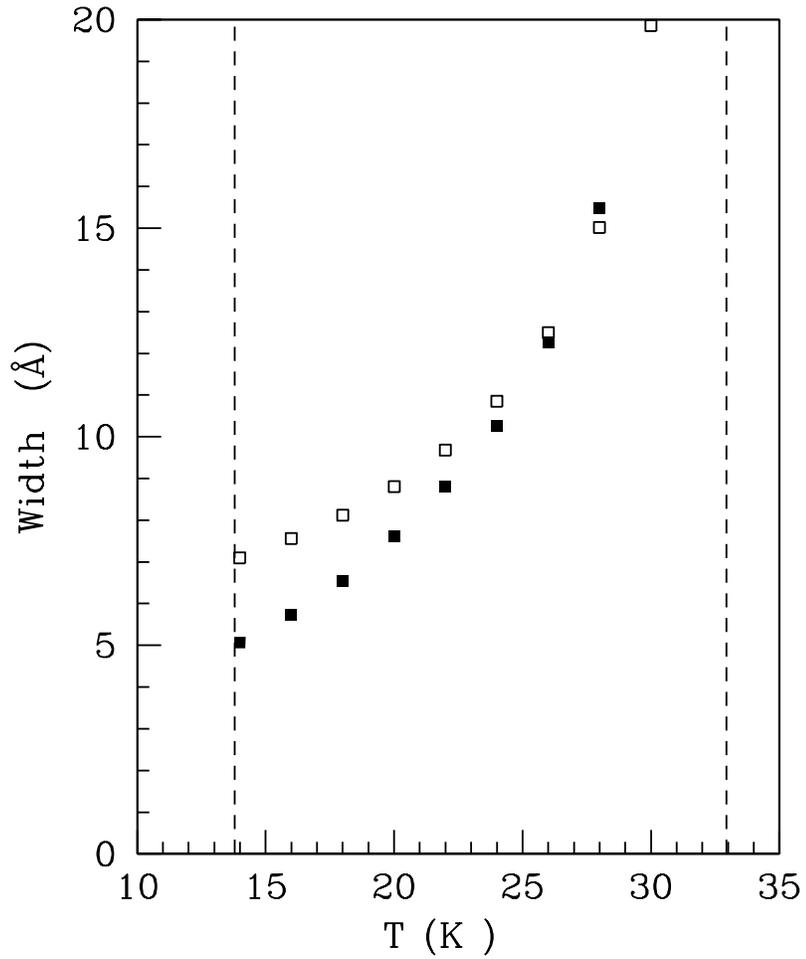}}
\caption{Thickess of the liquid-vapor interface. Open squares: results of the
zero-range DF. Full squares: results of the finite-range DF. 
The vertical lines indicate the triple and critical temperatures.
}
\label{fig4}
\end{figure}

We show in Fig. \ref{fig4} the results obtained for the thickness
of the liquid-vapor interface. We define the thickness as the distance
across the interface between the points at which the fluid density
changes from 0.9 to 0.1 times that of the liquid at saturation for
the corresponding $p$ and $T$ values.
Open squares represent the values obtained using the zero-range DF, and
full squares those obtained using the finite-range DF. The latter are
our prediction for the thickness of the interface.
Similarly to what occurs for liquid $^4$He,\cite{Dal95,Anc00} 
the thickness of the liquid-vapor interface obtained from the
finite-range DF at low temperatures is about 2 \AA{} smaller 
than that obtained from the zero-range DF.

\subsection{Wetting properties: the pH$_2$/Rb system}

The existence and nature of the wetting transition\cite{Cah77} at a
temperature above the triple point $T_{t}$ (and below the critical
temperature) is known to be the result of a delicate balance between the
interatomic potential acting among the atoms in the liquid and the
atom-substrate interaction potential. When the latter
dominates, the fluid tends to wet the surface. 

When a first order transition from partial to complete wetting
occurs at a temperature $T_{w}$ above the triple-point (and below the bulk
critical temperature), then a locus of first-order surface phase transitions 
must extend away from the liquid-vapor coexistence curve, on the
vapor side. At $T<T_{w}$ the thickness of the adsorbed liquid film
increases continuously with pressure, but the film remains microscopically
thin up to the coexistence pressure $p_{sat}(T)$, and becomes infinitely
(macroscopically) thick just above it. 
At temperatures $T_{w}<T<T_{pw}^{c}$,
$T_{pw}^{c}$ being the prewetting critical temperature, the thin film
grows as the pressure is increased until a transition pressure 
$p_{tr}(T)<p_{sat}(T)$ is reached. At this pressure a thin film is in
equilibrium with a thicker one, and a jump in coverage occurs as $p$
increases through $p_{tr}(T)$. At still higher pressures this first order
transition is followed by a continuous growth of the thick film which
becomes infinitely thick at $p_{sat}(T)$.\cite{Che93,Bon01}

Wetting transitions have been observed for quantum fluids (helium and
hydrogen) on alkali metal surfaces. In particular,
H$_2$ on Rb surface shows a wetting transition.\cite{Mis94}
This system has been addressed by
finite temperature quantum simulations in Ref. \onlinecite{Shi03},
where a first-order wetting transition was indeed found to occur.
The value found for the wetting temperature, $T_W=25.5 \pm 0.5$ K
was higher than the experimental value $T_W=19.1$ K.\cite{Mis94} 
The disagreement has been attributed to the fact that the
interaction potential used to simulate the fluid-surface interaction
was probably too weakly attractive. We are not concerned here in
reproducing the experimental value for $T_W$ but rather to compare
the prediction of our finite-range DF approach against an accurate benchmark,
{\it i.e.} the virtually exact Quantum Monte Carlo results
reported in Ref. \onlinecite{Shi03}.
The external potential $V_{s}(z)$ ($z$ is the coordinate normal to the surface
plane) used in Ref. \onlinecite{Shi03} to describe the
H$_2$/Rb surface interaction is a semi-{\it ab initio} one.\cite{Chi98}
In order to make a proper comparison with the results of 
Ref. \onlinecite{Shi03} we have used the same atom-surface
potential employed there, and calculate within finite-range 
DF theory the wetting temperature for the pH$_2$/Rb system.

The free energy finite-range functional for pH$_2$ 
described  in Sec. 2.2.2 must be augmented by a term
describing the interaction (per unit surface) of the pH$_2$ fluid with the Rb surface, 
$\int dz \rho (z) V_s(z)$, and
the equilibrium density profile $\rho (z)$ of the fluid adsorbed on the
surface is determined by direct minimization of the density functional
with respect to density variations, which yields an equation similar to
Eq. (\ref{eq:kseq}) with an extra potential term arising from
$V_s(z)$.
Details on the procedure can be found {\it e.g.,} in Ref. \onlinecite{Anc09}
and references therein.

\begin{figure}[t]
\centerline{\includegraphics[width=14cm,clip]{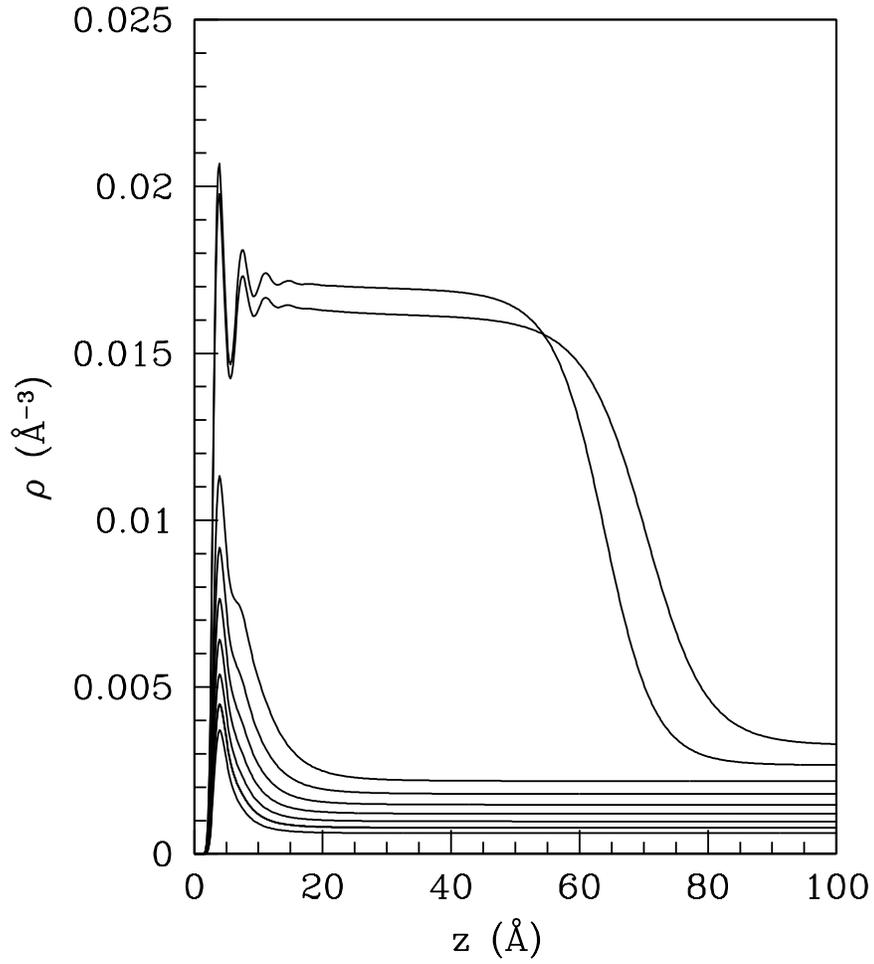}}
\caption{Density profiles of pH$_{2}$ on a planar Rb surface at different
temperatures. From bottom to top, in the asymptotic vapor regime:
low density profiles, $T=22$, 23, 24, 25, 26, 27, and 28 K;
high density profiles, $T=29$ and 30 K.
}
\label{fig5}
\end{figure}

We show in Fig. \ref{fig5} a set of calculated density profiles for
different temperatures. 
We fix in our calculations the pressure (equivalently, the chemical potential) 
just below the bulk liquid saturation pressure, and increase the temperature.
The abrupt (first order) change in the amount of liquid adsorbed on the
surface shown in Fig. \ref{fig5} when $T$ is just above $T=28$ K
is a clear signature of the pre-wetting transition very close
to saturation. We thus take as our prediction for $T_W$ the 
value $T_W=28.5 \pm 0.5$ K.
This value compares 
well
with the one found
by 
Quantum Monte Carlo calculations.\cite{Shi03}

\section{Cavitation}

Another application of the density functionals 
described
in Sec. 2 is the study of thermal homogeneous
cavitation in liquid pH$_2$. It can be straightforwardly
extended to the case of nucleation of pH$_2$ droplets in a supersaturated
parahydrogen vapor. From the experience gathered in similar $^4$He studies,
it is enough to use the zero-range DF for this application.
We refer the reader {\it e.g.}, to Refs.
\onlinecite{Xio91,Jez93,Pet94,Bar02,Bal02} and references therein
for a general discussion of similar work carried out for liquid helium.

It is well know that phase transitions take place in the coexistence
region of different phases of homogeneous systems, and that it does not
always happen under equilibrium conditions. Indeed, as the coexisting
phase forms, the free energy of the system is lowered, but the original
phase can be held in a metastable state close to the equilibrium
transition point. The metastable state is separated from the stable
state by a thermodynamic barrier that can be overcome by statistical
fluctuations. 

Cavitation consists in the formation of bubbles in a liquid held at a
pressure below the saturation vapor pressure at the given temperature.
In this case, cavitation is a thermally driven process. The formation
of the vapor phase from the metastable liquid proceeds through the
formation of a critical bubble, whose expansion cannot be halted as its 
energy equals the thermodynamic barrier height. A very simple estimate of 
the radius $R_c$ of this critical bubble is provided by the thin
wall model, which consists in treating the bubble as a spherical cavity
of sharp surface and radius $R$ filled with  vapor at the saturated
vapor pressure, $p_{svp}$. The barrier for creating a bubble of radius
$R$ is the result of a balance between surface and volume energy terms

\begin{equation}
U(R)= 4 \pi \gamma R^2 - \frac{4 \pi}{3} R^3 |\Delta p| \; ,
\label{cav1}
\end{equation}
where  $|\Delta p|$ is the difference between the applied pressure and 
$p_{svp}$, and can be positive and negative as well. This barrier
has thus a maximum $U_{max}= 16 \pi \gamma^3/(3 |\Delta p|^2)$ at a critical
radius $R_c= 2 \gamma/|\Delta p|$. Taking $\gamma= 2.96$
dyne/cm at $T$= 14 K,\cite{Mcc81}  and
$|\Delta p| \sim 78$ bar (the pH$_2$ spinodal pressure at $T= 14$ K),
one gets $R_c \sim 7.5$ \AA{}. 

The cavitation rate $J$, {\it i.e.,} the number of critical bubbles
formed in the liquid per unit time and volume is given by

\begin{equation}
J= J_{0T} \, \exp(-U_{max}/T) \; ,
\label{cav2}
\end{equation}
and the prefactor $J_{0T}$ depends on the dynamics of the cavitation process. 
A precise knowledge of the prefactor is not crucial, since at given $T$,
the exponential term in Eq. (\ref{cav2}) dramatically varies with $p$
in the pressure range of interest, thus making the determination of 
$p_h$ rather insensitive to its actual value.\cite{Jez93} We shall 
estimate $J_{0T}$ as an attempt frequency per 
unit volume, namely $J_{0T}= T/(h V_c)$, where $h$ is the Planck constant
and $V_c$ is the volume of the critical bubble.

To determine the homogeneous cavitation pressure within DF theory, one has 
first to determine the critical bubble for $p$ and $T$ values
corresponding to points lying within the liquid-vapor phase equilibrium region
where bubbles may appear. This is done by solving the EL  equation
arising from the functional variation of the grand potential with respect to the
pH$_2$ particle density. The grand potential density is $\omega(\rho, T) =
f(\rho, T) - \mu \rho$, where 
$f(\rho, T)$ is the free energy per unit volume including surface terms arising
from the bubble surface and kinetic energy contributions already discussed
in Sec. 2.2.1, and
$\mu$ is the chemical potential of the bulk liquid. The EL equation is solved
imposing the physical conditions that $\rho'(r=0)= 0$ and 
$\rho(r \rightarrow \infty )= \rho_l$, where $\rho_l$ is the density of the metastable
homogeneous liquid.

The nucleation barrier $U_{max}$ is obtained by subtracting the grand potential of
the critical bubble from that of the homogeneous metastable liquid at the given 
$p$ and $T$ 

\begin{equation}
U_{max}= \int d\mathbf{r} \, [f(\rho, T) - f_{l}(\rho_l, T) - \mu
(\rho- \rho_l)] \; ,
\label{cav3}
\end{equation}
where $f_{l}(\rho_l, T)$ has no density gradient terms as it corresponds to the
metastable homogeneous liquid. The above expression embraces the two limiting physical
situations of a barrierless process when the system approaches the
spinodal point, and of an infinite barrier process when the system approaches
the saturation line.
While the latter limit is well reproduced by the thin wall model, the former is 
not.\cite{Jez93,Bar02}

\begin{figure}[t]
\centerline{\includegraphics[width=14cm]{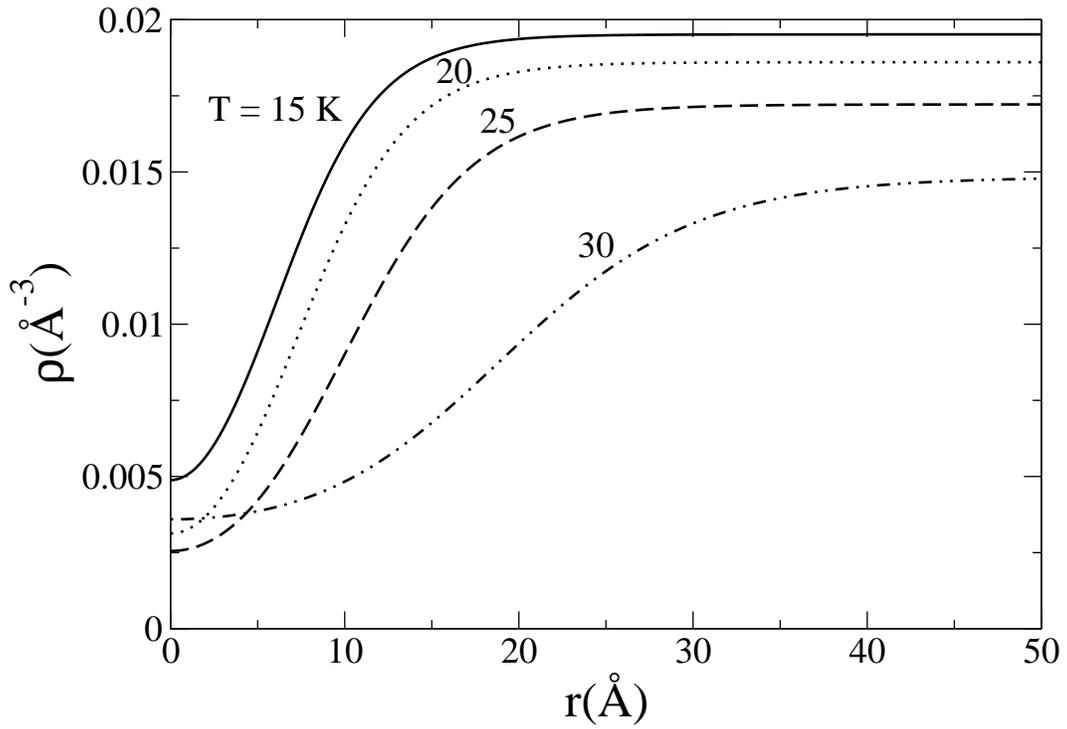}}
\caption{Density profiles of critical bubbles corresponding to several $T$
values obtained using the zero-range DF.
For each temperature, we have chosen the pressure
corresponding to the homogeneous cavitation value.}
\label{fig6}
\end{figure}

Figure \ref{fig6} shows several density profiles for critical bubbles at
selected $T$ values. For each temperature, we have chosen the pressure
corresponding to the homogeneous cavitation value, see below.
The radius of the critical bubble for $T=15$ K is similar to that obtained
within the thin wall model at $T=14$ K.

\begin{figure}[t]
\centerline{\includegraphics[width=14cm]{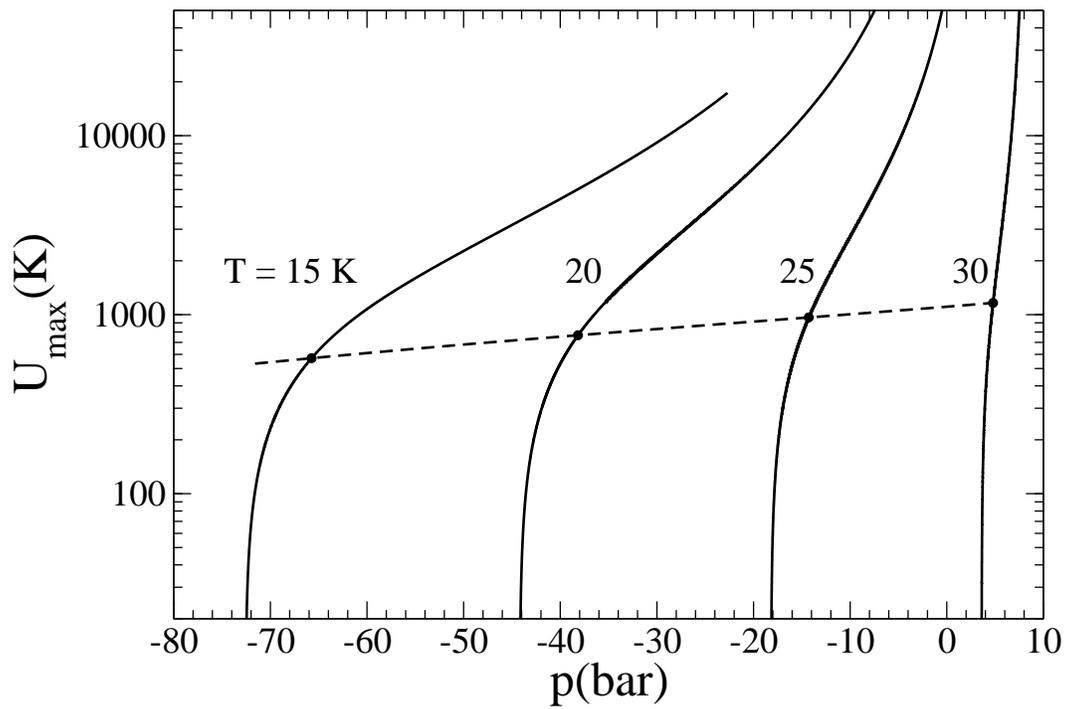}}
\caption{Solid lines: barrier height $U_{max}$ of critical bubbles for
selected $T$ values (K) as a function of $p$ (bar).
The dashed line represents the barrier height at 
$p=p_h$ for the chosen value of $T$.
}
\label{fig7}
\end{figure}

Figure \ref{fig7} shows the barrier height of the critical bubble for
selected $T$ values as a function of $p$ (solid lines).
The divergence of $U_{max}$ as $p$ approaches the saturation line,
as well as its disappearance at the spinodal line are clearly visible in the figure.

At a given temperature, there is an appreciable probability of cavitation occuring when $p$ 
reaches a value such that   

\begin{equation}
1= (V \times t)_e \, J_{0T} \, \exp(-U_{max}/T) \; ,
\label{cav4}
\end{equation}
where $(V \times t)_e$ is the experimental volume times the experimental time.
This defines the homogeneous cavitation pressure $p_h$,
an intrinsic property of the liquid 
that is determined by solving the above algebraic equation.
To solve Eq. (\ref{cav4}) one needs some experimental information.
If the experiment were carried out as for liquid $^4$He,\cite{Cau01}
$(V \times t)_e \sim 2 \times 10^{-16}$ cm$^3$ s. Lacking of a better
choice, this is the value we shall adopt here for parahydrogen.

\begin{figure}[t]
\centerline{\includegraphics[width=14cm]{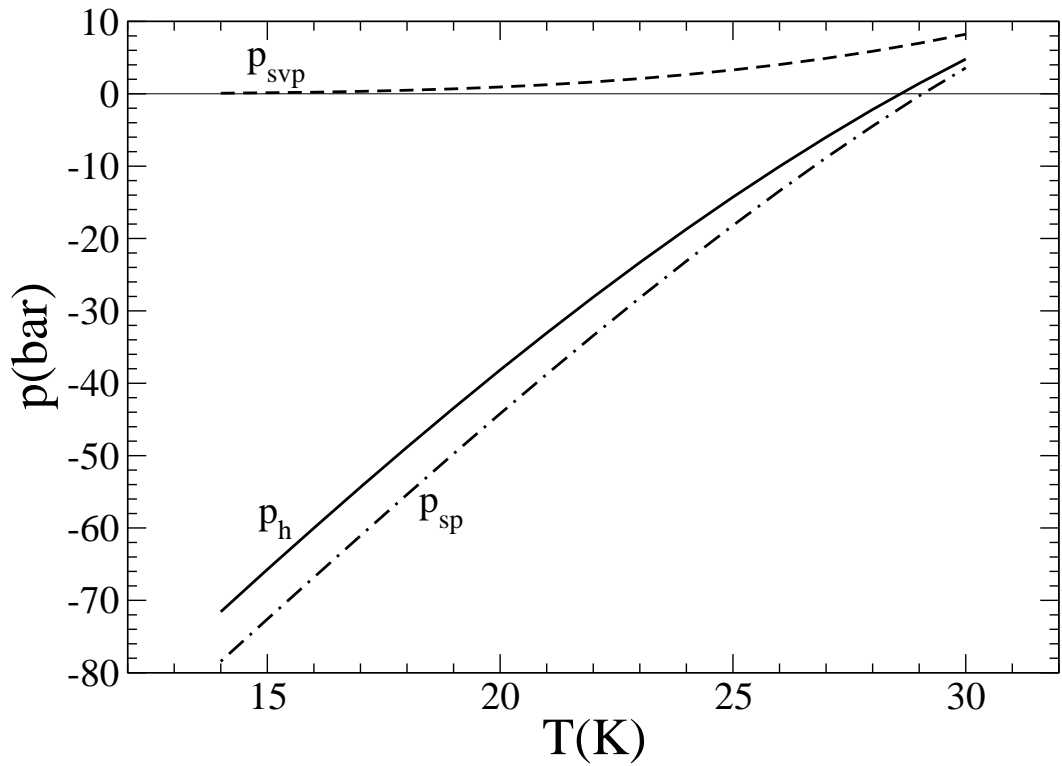}}
\caption{Homogeneous cavitation pressure $p_h(T) $ as a function
of temperature. Also drawn are the spinodal line $p_{sp}(T)$ and 
saturation vapor pressure line $p_{svp}(T)$. 
}
\label{fig8}
\end{figure}

Figure \ref{fig8} shows the homogeneous cavitation pressure as a function 
of $T$. Also shown is the spinodal line $p_{sp}(T)$ and the saturation
vapor pressure line $p_{svp}(T)$. The three lines merge at the critical 
point. It can be seen that $p_h$ is negative up to $T \sim 29$ K,
meaning that a large tensile strength may be needed to produce cavitation.
Notice that, as in $^4$He,\cite{Xio91,Jez93} the homogeneous cavitation 
pressure is 
just above
the spinodal pressure.

Finally, we mention that the dashed line in Fig. \ref{fig7} represents the
barrier height at $p_h$ for the chosen $T$ value. It can be seen that
these heights are in the 500-1000 K range for temperatures
$14 \leq T \leq 30$ K. 

\section{Summary and outlook}

In this work we have 
proposed
a density functional that can be used to
study the properties of parahydrogen in the liquid-vapor phase-diagram
region and its neighborhood. We have used its zero-range version 
to address thermal homogeneous cavitation in the metastable liquid,
and its finite-range version to predict the thickness of the liquid-vapor
interface. We have shown that this finite-range DF predicts the surface
tension of the liquid-vapor interface and the wetting properties of pH$_2$ on 
planar Rb surfaces in satisfactory agreement with experiments
and microscopic calculations, respectively.

Besides thermal cavitation, the existence of a finite temperature,
finite-range DF makes it possible to improve the theoretical description 
of phenomena that hardly could be described using more fundamental approaches,
as the study of electron bubbles in pH$_2$.\cite{Lev94,Ber03} Large pure or
doped parahydrogen drops at finite temperature could also be addressed provided 
the impurity-pH$_2$ interaction is known, as well as extending the study of wetting
to other substrates and/or geometries.

The present DF may be of interest for other applications,
such as the calculations of the excitation
spectra of pure or doped pH$_2$ drops, as well as their static or dynamic responses. 
The extrapolation of the DF to temperatures near $T=0$ 
still
remains an open question. 
We plan to study some of these phenomena in the future.

\section*{Acknowledgments}
This work has been performed under Grants No. FIS2008-00421/FIS and FIS2007-60133 from 
DGI, Spain (FEDER), and Grant 2009SGR01289 from Generalitat de Catalunya.

\newpage
\onecolumn

\begin{table}[t]
\caption{DFpH2 set of parameters $a_i(T)$ (in K  \AA$^{3i}$) entering the correlation 
energy density functional $f_c$. The parameter $h(T)$ (in \AA) related to the 
finite-range DF is given in the last column.}
\label{coefs}
\begin{center}
\begin{tabular}{c|cccc|c}
\hline
$T$ & $a_1/10^3$ & $a_2/10^5$ & $a_3/10^7$ & $a_4/10^8$ & $h$ \\ \hline
14 & -5.39621105 & 2.25023742 & -1.20053135 & 2.79594232 & 2.98886735   \\  
15 & -5.28361552 & 2.15103496 & -1.14667549 & 2.70797035 & 2.93284836   \\
16 & -5.18646074 & 2.07504193 & -1.10362936 & 2.63656153 &  2.90255249 \\
17 & -5.10162915 & 2.01565556 & -1.06770278 & 2.57547938 &  2.88158016 \\
18 & -5.02544156 & 1.96727851 & -1.03616377 & 2.52033226 &  2.86543169 \\
19 & -4.95500617 & 1.92571884 & -1.00701996 & 2.46790257 &  2.85212913 \\
20 & -4.88844024 & 1.88827147 & -0.978964063 & 2.41596399&  2.84067446 \\
21 & -4.82469889 & 1.85328416 & -0.951137931 & 2.36290741 &  2.83053490  \\
22 & -4.76330170 & 1.81966306 & -0.922799978 & 2.30707709 &  2.82141283 \\
23 & -4.70410834 & 1.78695682 & -0.893590816 & 2.24744238 & 2.81313315  \\
24 & -4.64712668 & 1.75458070 & -0.862880207 & 2.18218726 &  2.80558079  \\
25 & -4.59255499 & 1.72283140 & -0.830703205 & 2.11075548 &  2.79868923  \\
26 & -4.54035421 & 1.69068884 & -0.795990139 & 2.03000386 & 2.79237811  \\
27 & -4.49046505 & 1.65741296 & -0.757823418 & 1.93673005 &  2.78657987 \\
28 & -4.44267216 & 1.62195349 & -0.714989528 & 1.82655914 &  2.78122146 \\
29 & -4.39598018 & 1.58109948 & -0.664408250  & 1.68988952 &  2.77615776   \\
30 & -4.34840634 & 1.52965533 & -0.601172851 & 1.51084983 &  2.77116004  \\
31 & -4.29478440 & 1.45544413 & -0.514224892 & 1.25423884 &   2.76570856  \\
32 & -4.21484337 & 1.31639928 & -0.367772398 & 0.809972555&  2.75790923 \\
  \hline
  \end{tabular}
  \end{center}
  \end{table}

\newpage
\pagebreak

\begin{thebibliography}{99}

\bibitem{Str87} Stringari, S.; Treiner, J.
\textit{Phys. Rev. B} \textbf{1987}, \textit{36}, 16;
\textit{J. Chem. Phys.} \textbf{1987}, \textit{87}, 5021.

\bibitem{Dal95} Dalfovo, F.; Lastri A.; Pricaupenko, L.; Stringari, S.; Treiner, J.
\textit{Phys. Rev. B} \textbf{1995}, \textit{52}, 1193.

\bibitem{Bar06} M. Barranco, M.; Guardiola, R.; Hern\'andez, S.; Mayol, R.;
Pi, M.
\textit{J. Low Temp. Phys.} \textbf{2006},  \textit{142}, 1.

\bibitem{Gin72} Ginzburg, V.L.;  Sobyanin, A.A.
\textit{Sov. Phys. JETP Lett.} \textbf{1972}, \textit{15}, 242. 

\bibitem{Mar83} Maris, H.J.; Seidel, G.M.; Huber, T.E. 
\textit{J. Low Temp. Phys.} \textbf{1983},  \textit{51}, 471.

\bibitem{Alo05} Alonso, J.A. 
\textit{Structure and Properties of Atomic Clusters}. Imperial College Press:  
London, (2005). 

\bibitem{Sin91} Sindzingre, Ph.; Ceperley, D.M.; Klein, M.L. 
\textit{Phys. Rev. Lett.} \textbf{1991}, \textit{67}, 1871.

\bibitem{Gre00} Grebenev, S.; Sartakov, B.; Toennies, J.P.; Vilesov A.F. 
\textit{Science} \textbf{2000}, \textit{289}, 1532.

\bibitem{Tej04} Tejeda, G.; Fern\'andez, J.M.; Montero, S.; Blume, D.; Toennies, J.P. 
\textit{Phys. Rev. Lett.}  \textbf{2004}, \textit{92}, 223401.

\bibitem{Mon09} Montero, S.; Morilla, J.H.; Tejeda, G.; Fern\'andez, J.M. 
\textit{Eur. Phys. J. D} \textbf{2009}, \textit{52}, 31.

\bibitem{War10} Warnecke, S.; Sevryuk, M.B.; Ceperley, D.M.; Toennies, J.P.; 
Guardiola, R.; Navarro J. \textit{Eur. Phys. J. D} \textbf{2010}, \textit{56}, 353.

\bibitem{Alo10} Alonso, J.A.; Mart\'{\i}nez, J.I. 
\textit{Handbook of Nanophysics. Vol 2: Clusters and Fullerenes}, Sattler, K.D., Ed.; 
Taylor and Francis: Boca Raton, 2010. 

\bibitem{Gua10} Navarro, J.; Guardiola, R.
\textit{Int. J. Quantum Chem} \textbf{2011}, \textit{111}, 463.

\bibitem{Gor97} Gordillo, M.C.; Ceperley, D.M.
\textit{Phys. Rev. Lett.} \textbf{1997}, \textit{97}, 3010.

\bibitem{Mis94} Mistura, G.; Lee, H.C.; Chan, M.H.W.
\textit{J. Low Temp. Phys.} \textbf{1994}, \textit{96}, 221.

\bibitem{Ros98} Ross, D.; Taborek, T.; Rutledge, J.E.
\textit{Phys. Rev. B} \textbf{1998}, \textit{58}, 4247

\bibitem{Eva89} Evans, R.
\textit{Liquids at interfaces}, (Editors
J. Charvolin, J. F. Joanny, and J. Zinn-Justin).
Elsevier, (1989).

\bibitem{Bon04} Boninsegni, M.; Szybisz L.
\textit{J. Low Temp. Phys.} \textbf{2004}, \textit{134}, 315.

\bibitem{Gui92} Guirao, A.; Centelles, M.; Barranco, M.; Pi, M.; Polls, A.; Vi\~nas, X.
\textit{J. Phys.: Condensed Matter} \textbf{1992}, \textit{4}, 667.

\bibitem{Anc00} Ancilotto, F.; Faccin, F.; Toigo, F.
\textit{Phys. Rev. B} \textbf{2000}, \textit{62}, 17035.

\bibitem{Hua87} Huang, K.
\textit{Statistical Mechanics}, J. Wiley: New York, 1987, 2nd. edition. 

\bibitem{Mcc81} McCarthy, R.D.; Hord, L.; Roder, H.M.
\textit{National Bureau of Standards} \textbf{1981}, \textit{Monograph num. 168}.

\bibitem{Lea09} Leachman, J.W.; Jacobsen, R.T.; Penoncello, S.G.; Lemmon, E.W.
\textit{J. Phys. Chem. Ref. Data} \textbf{2009}, \textit{38}, 721.

\bibitem{Bar90} Barranco, M.; Pi, M.; Polls, A.; Vi\~nas, X.
\textit{J. Low Temp. Phys.} \textbf{1990}, \textit{80}, 77.

\bibitem{Sil78} Silvera, I.F.; Goldman, V.V.
\textit{J. Chem. Phys.} \textbf{1978}, \textit{69}, 4209.

\bibitem{Zha04} Zhao, X.; Johnson, J.K.; Rasmussen, C.E.
\textit{J. Chem. Phys.} \textbf{2004}, \textit{120}, 8707.

\bibitem{Cah77} Cahn, J.W.
\textit{J. Chem. Phys.} \textbf{1977}, \textit{66}, 3667.

\bibitem{Che93} 
Cheng, E.; Cole, M.W.;  Dupont-Roc, J.; Saam, W.F.; Treiner, J.
\textit{Rev. Mod. Phys.} \textbf{1993}, \textit{65}, 557.

\bibitem{Bon01}  Bonn, D.; Ross, D.
\textit{Rep. Prog. Phys.} \textbf{2001}, \textit{64}, 1085.

\bibitem{Shi03} Shi, W.; Johnson, J.K.;  Cole, M.W.
\textit{Phys. Rev. B} \textbf{2003}, \textit{68}, 125401.

\bibitem{Chi98} Chizmeshya, A.; Cole, M.W.; Zaremba, E.
\textit{J. Low Temp. Phys.} \textbf{1998}, \textit{110}, 677.

\bibitem{Anc09} Ancilotto, F.; Barranco, M.B.; Hern\'andez, E.S.;
Pi, M.
\textit{J. Low Temp. Phys.} \textbf{2009}, \textit{157}, 174.

\bibitem{Xio91} Xiong, Q.; Maris, H.J.
\textit{J. Low Temp. Phys.} \textbf{1991}, \textit{82}, 105.

\bibitem{Jez93} Jezek, D.M.; Guilleumas, M.; Pi, M.; Barranco, M.; 
Navarro, J. 
\textit{Phys. Rev. B} \textbf{1993}, \textit{48}, 16582.

\bibitem{Pet94} Pettersen, M.S.; Balibar, S.; Maris, H.J.
\textit{Phys. Rev. B} \textbf{1994}, \textit{49}, 12062.

\bibitem{Bar02} Barranco, M.; Guilleumas, M.; Pi, M.; Jezek, D.M.
\textit{Microscopic Approaches to Quantum Liquids in Confined
Geometries}, E. Krotscheck and J. Navarro, editors.
World Scientific: Singapore, (2002), p. 319.

\bibitem{Bal02} Balibar, S.
\textit{J. Low Temp. Phys.} \textbf{2002}, \textit{129}, 363.

\bibitem{Cau01} Caupin, S.; Balibar, S.
\textit{Phys. Rev. B} \textbf{2001}, \textit{64}, 064507.

\bibitem{Lev94} Levchenko, A.A.; Mezhov-Deglin, L.P.
\textit{JETP Lett.} \textbf{1994}, \textit{60}, 470.

\bibitem{Ber03} Berezhnov, A.V.; Khrapak, A.G.; Illenberger, E.;
Schmidt, W.F.
\textit{High Temperature} \textbf{2003}, \textit{41}, 425.

\end{thebibliography}
\end{document}